%% file: maintext__for_preprint_post_.tex
\documentclass[poms,final,nonblindrev]{poms1_V1} 

\OneAndAHalfSpacedXI 

\usepackage{graphicx}
\usepackage{subfigure, epsfig}
\usepackage{natbib}

 \bibpunct[, ]{(}{)}{,}{a}{}{,}%
 \def\bibfont{\small}%
\TheoremsNumberedThrough     
\ECRepeatTheorems

\EquationsNumberedThrough    

\newtheorem{subhypothesis}{Hypothesis}[hypothesis]

\usepackage{booktabs}
\usepackage{rotating}
\usepackage{lscape}
\usepackage{longtable}
\usepackage{siunitx}

\begin{document}


\RUNAUTHOR{Sun, Jia, and Oliva }

\RUNTITLE{The Impact of Carbon Targets}

\TITLE{The Impact of Carbon Targets on Firms' Carbon Performance}

\ARTICLEAUTHORS{%
\AUTHOR{Xichen Sun}
\AFF{Department of Information Systems and Operations Management, Tilburg University, \EMAIL{xichen.sun@tilburguniversity.edu}} 
\AUTHOR{Xingzhi Jia}
\AFF{Department of Management, Renmin University of China, \EMAIL{jiaxingzhi@rmbs.ruc.edu.cn}} 
\AUTHOR{Rogelio Oliva}
\AFF{Department of Information and Operations Management, Texas A\&M University \EMAIL{roliva@tamu.edu}}
} 

\ABSTRACT{Although goal-setting theory predicts that a challenging and specific goal can improve performance, the evidence regarding the effectiveness of an environmental goal in organizations is mixed. Using a panel data set consisting over 700 global firms across multiple industries, we apply a Difference-in-Differences (DiD) strategy and empirically analyze the effect of having science-based carbon emissions targets (SBTs) on firms’ carbon performance. The findings in this study contribute to the debate regarding the necessity of setting SBTs.

}%


\maketitle

\section{Introduction}\label{sec:intro}
To combat climate change and prevent catastrophic consequences, leaders from 196 parties globally reached the Paris Agreement in 2015 and set a long-term objective to limit global warming to well below 2 degrees Celsius compared to the pre-industrial level \citep{UNFCCC2022paris}. Answering the call to this commitment, an increasing number of firms have adopted carbon reduction targets. These targets are reviewed and certified as science-based targets (SBTs) by a non-profit organization, the Science-Based Target initiative (SBTi)\footnote{The certification ensures the target is sufficiently ambitious such that it aligns with the goal to limit global warming well below 2 degrees Celsius compared to the pre-industrial level. However, it does not include the certification of means to achieve SBTs.}. The SBTi works with firms to set up SBTs that reflect their fair share of carbon emissions reduction across sectors. As of January 2023, over 4,000 firms, accounting for more than a third of global market capitalization, have either adopted or committed to adopting SBTs \citep{SBTi2022progress}.  

Despite their prevalence, empirical evidence regarding the effectiveness of SBTs is limited. Goal-setting theory predicts that a challenging and specific goal can improve performance \citep{locke2002building}. Consistent with the theory, the SBTi, the biggest advocate of SBTs, reported that firms adopting SBTs achieved greater carbon emissions reduction than those without SBTs during the 2015-2021 period \citep{SBTi2022progress}. However, research has found that many firms with SBTs have a track record of superior carbon performance prior to adoption \citep{freiberg2021science}. \cite{giesekam2021science} noted that many SBTs were achieved within the year of target approval, raising the question of whether SBT achievement can be attributed to \textit{``inspired rapid action or rewarded historical achievement"} (p.14). Like much other sustainability propaganda, STBs can fall prey to \textit{greenwashing} firms that implement sustainable practices only to manipulate stakeholder perceptions and do not make substantive efforts. As a result, some opponents called SBTs a \textit{costly distraction} \citep[][p.931]{trexler2015science}. Whether an SBT can truly help accelerate firms’ decarbonization progress, as suggested by the goal-setting theory, remains unclear. 

Even if having an SBT can contribute to the progression of carbon reduction, the effectiveness of SBTs may vary from firm to firm. As firms with superior historical carbon performance typically have established environmental management systems (e.g. ISO 14001) and substantial experience in carbon reduction practices, it gives them an advantage against peers when on the course of achieving SBTs. However, these firms may find it increasingly challenging to keep up with the reduction rate required by SBTs over the long term as they move beyond the low-hanging fruit and start to face increasing costs and diminishing returns \citep{wu2011balancing}. In contrast, firms with inferior historical carbon performance, despite their lack of experience, may achieve greater carbon reductions with the guidance of SBTs due to their potential for improvement. The question becomes: who will benefit more from having an SBT, the best-in-class firms or the worst-in-class ones?
 
From an operations perspective, adopting an SBT can create tensions between carbon reduction efforts and other non-carbon sustainability projects. After adopting the target, the external pressure to adhere to the SBT can drive managers to reallocate their attention and thereby impede the progress of other ongoing sustainability efforts \citep{dhanorkar2018promoting}. Further, due to limited resource availability, when firms with ambitious carbon targets start to invest more in carbon reduction initiatives, the investment in other programs can decrease. Finally, trade-offs can happen when managing multiple goals like carbon emissions and water usage simultaneously. American Electric Power Company, for example, reported its concern about increased water consumption due to the transition from coal-fired power plants to gas-based generation \citep{CDP2019water}. Therefore, it remains a question how the continuous effort to achieve an SBT may affect other non-carbon-related sustainability performances. 

Based on the discussion above, we intend to answer the following research questions: 1) Does adopting SBTs improve firms’ carbon performance? 2) How does a firm's relative carbon performance affect the effectiveness of SBTs? 3) Does adopting SBTs affect other sustainability-related performances? Using panel data provided by Refinitiv (formerly Thomson Reuters) and the SBTi and applying a Difference-in-Differences (DiD) identification strategy with Propensity Score Matching (PSM), we first analyze the impact of having SBTs on firms' carbon reduction progress. We find that a small portion of early SBT adopters managed to improve their carbon performance four to five years after adopting SBTs, but this does not always translate to an overall significant treatment effect for all treated firms. Next, we explore the performance difference between best- and worst-in-class firms after adopting SBTs. In the base model, we classify the 10\% of firms with the highest industry-year adjusted carbon emissions as the worst-in-class and the 10\% of firms with the lowest carbon emissions as the best-in-class. We find that these firms do not differ in carbon performance after adopting SBTs. In the additional analyses conducted using alternative cut-off points, we find that the best 20\% and 25\% firms reduced carbon emissions more aggressively after adopting SBTs. Finally, using Refinitiv's environmental and social pillar scores as proxies for firms' sustainability performance at two aggregate levels, we find that adopting SBTs does not affect firms' overall environmental and social sustainability performances. 

This study contributes to the debate about whether encouraging firms to adopt SBTs is an effective strategy or a costly distraction amid the combat against global warming. The findings in this study reveal that both the proponents and opponents of SBTs were partially right about their argument. That is, although firms with superior carbon performance are more likely to adopt SBTs, they do benefit from having one. In contrast, the worst-in-class firms' carbon performance does not improve after adopting SBTs. They either have not realized the benefits of SBTs due to the delay in technological and structural development essential for reducing carbon emissions or are simply not committed to making substantive efforts for goal achievement. As most of the treated firms adopted SBTs in 2020 and 2021, it is too early to make a definite statement regarding the effectiveness of SBTs. Nevertheless, this finding sends an important message to SBT proponents that having SBTs may not guarantee carbon emissions reduction. 

Contributing to the sustainable operations literature, our analysis reveals that a resource demanding program like SBTs does not necessarily create distractions against other non-carbon sustainability metrics at an aggregate level. Our interviews with senior managers at Flex Ltd. further confirmed that firms with well-established structures for sustainability experience fewer resource conflicts among programs. However, as the SBT program is still young and not all firms with SBTs may be making efforts to achieve them, we need more data to confirm this finding. We advise managers and social planners to interpret our findings with caution.

The rest of the paper is structured as follows. In \S \ref{sec:litreview}, we review the literature and link theories to our hypotheses, and \S \ref{sec:method} describes the data, measures, and model specifications. Results are presented in \S \ref{sec:result}, followed by a discussion on the implications of findings in \S \ref{sec:discussion}. We conclude with limitations and future research opportunities in \S \ref{sec:conclusion}.
\section{Theory and Hypotheses}\label{sec:litreview}
\subsection{Goal Setting in Organizations}\label{subsec:goalsetting}
The goal-setting theory developed in industrial/organizational psychology prescribes that a specific, challenging goal is in general positively associated with task performance \citep[see][for an overview]{locke2002building}. Goals not only motivate individuals to exert greater effort with persistence but also direct effort and attention to goal-relevant actions like the discovery and use of task-relevant knowledge. The effectiveness of goals improves when individuals are committed \citep{klein1999goal}, can apply appropriate skills and strategies as task complexity increases \citep{durham1997effects}, and receive timely and relevant feedback \citep{latham1991self}. In the operations management literature, goal setting has been proven effective in service quality \citep{hays2001preliminary,hezarkhani2022toward}, quality improvement programs \citep{sterman1997unanticipated,linderman2006six}, shop floor performance \citep{doerr1996impact,de2011improving}, distribution center performance \citep{doerr2013performance}, and energy-saving behavior in household and industrial settings \citep{jung2021repairing,dhanorkar2021nudges}.

Despite its usefulness in enhancing individual and group performance, goals in organizations can add complexity when multiple entities are involved, making outcomes less predictable. First, the classic agency theory \citep{eisenhardt1989agency} suggests that individuals may have personal goals that conflict with organizational goals. Organizations' long-term successes, for example, may not be managers' first priority if they are compensated by short-term performance, leading to managerial short-termism that can be detrimental for firms \citep{marginson2008exploring}. Further, goal conflicts also exist among functional departments. Departments with misaligned incentives can work against each other and create inefficiencies that slow down the goal-achieving progress \citep{oliva2009managing}. Third, goal conflicts can occur between players along supply chains, commonly observed in buyer-supplier relationships \citep{kanda2008supply}. These conflicts, if not coordinated, could undermine the mechanisms through which goals lead to greater and highly focused effort. The coordination itself, however, is a non-trivial task that can require skills and capabilities that are not possessed by participants, further influencing the effectiveness of a goal. Thus, it is not surprising that research using firms as the unit of analysis has discovered side effects \citep{ordonez2009goals} and limiting conditions \citep{latham1975review} that reveal mixed evidence of goal effectiveness in organizations \citep{arnold2015target}.

\subsection{Carbon Targets and Performance}\label{subsec:envtargets}
Facing increasing pressure to be environmentally and socially responsible, more and more firms have started to adopt and disclose responsible practices \citep{villena2020institutional}. Firms set up carbon targets for various reasons. Some do so as a response to stakeholder pressures (i.e., legitimacy), while others see opportunities to differentiate themselves from competitors and improve their image in the market \citep[i.e., signaling;][]{pinkse2013emergence}. Carbon targets vary in terms of scope (direct, indirect, or supply chain), scale (reduction percentage against the base year), coverage (the percentage of total carbon emissions covered by the target), measure (carbon intensity vs. absolute emissions level), and duration (number of years needed for target completion). Thus, firms tend to set up carbon targets that fit their own position and purpose. For example, Apple announced its plan to achieve carbon neutrality across its entire business by 2030, and BP is aiming to reach net-zero emissions by 2050. While it appears Apple is more ambitious in carbon reduction given its shorter time frame, BP's historical carbon emissions level was much higher than Apple's. It is not immediately clear which firm has a higher target. In the 2022 United Nations Climate Change Conference (COP27), the UN chief accused some net-zero carbon targets that exclude core products and activities polluting the planet of being a "toxic cover-up" and urged firms to "review their promises" to align with the Paris Agreement \citep{un2022cop27}. 

The various dimensions of carbon targets used in different firms make it challenging to assess their effectiveness. Several researchers investigate the influence of carbon targets on environmental performance, trying to differentiate substantive effort from greenwashing. \cite{ioannou2016effect} use target completion as a proxy for carbon performance and find that firms setting more difficult targets are associated with a higher percentage of completed targets. \cite{dahlmann2019managing} classify carbon targets as having substantive intentions if they lead to carbon emissions reduction. They find that (1) absolute carbon emissions targets, (2) more ambitious emissions reductions, and (3) a longer target time frame are associated with significant carbon emissions reduction. 

These findings, although informative, do not fully address the question about the effectiveness of carbon targets due to the selection bias of different carbon targets' dimensions and the potential trade-offs between these dimensions. The choice of carbon targets, as discussed above, is a function of the firm's current position and what it intends to achieve with the target. A firm with superior historical environmental performance, for example, may adopt an ambitious carbon target that aligns with their projected carbon footprint trajectory. Thus, the carbon reduction that occurs afterwards cannot be fully attributed to having a carbon target. For the same reason, completing a target does not necessarily mean better carbon performance, as some firms choose targets that can be completed within a year, while others adopt long-term targets with no intention to complete them at all.  

\subsection{Science-Based Targets}\label{subsec:sbts}
The emergence of SBTs changes the way carbon targets are perceived. SBTs provide a universal standard that is comparable among firms across industries. The SBTi offers guidance for firms to set SBTs aligned with the science behind the Paris Agreement and evaluates and certifies targets as science-based once approved. The universal standard alleviates the concern that firms could choose convenient targets based on their needs\footnote{The SBTi recommends two methods for setting SBTs: the Absolute Contraction Approach (ACA) or the Sectoral Decarbonization Approach (SDA). It also develops some sector-specific methods as variants of ACA or SDA. Although the different methods used may still impact target difficulties, the extent to which one target is easier than another to achieve is lesser. For more information, please refer to \url{https://sciencebasedtargets.org/sectors}}. The SBTi encourages and sets up additional rules for progress reporting to further reduce opportunism\footnote{For example, the SBTi prohibits the use of offsets as means to achieve SBTs. \url{https://sciencebasedtargets.org/resources/files/SBTi-Corporate-Manual.pdf}}. The decision on multiple dimensions of carbon targets now reduces to one: whether to adopt an SBT or not.

Although firms were made aware and started adopting SBTs only in the last few years, as more data became available, research on the drivers for and the effects of SBTs began to emerge. Taking a sample of firms with carbon targets, \cite{freiberg2021science} find that firms with a track record of setting and achieving ambitious carbon targets are more likely to adopt SBTs. Their results also suggest that adopting SBTs is accompanied by an increase in target difficulty, more investments in carbon projects, and higher carbon reduction and monetary savings from those projects. Conversely, in a similar study, \cite{bolton2021firm} find that overall firms committing to adopt SBTs (including those with SBTs not yet approved by SBTi) do not necessarily perform better than their counterparts in terms of scope 1 carbon emissions (i.e. direct greenhouse gas emissions from sources that are owned and controlled by the firm). Although this result can be partially attributed to the fact that firms with not-yet-approved SBTs are less committed to reducing carbon emissions, the mixed evidence in these studies does raise the question again: does adopting SBTs improve firms' carbon performance?

Different from arbitrary carbon targets set based on firms' internal standards, SBTs are ambitious and thus difficult to achieve \citep{freiberg2021science}. According to the goal-setting theory, difficult goals typically lead to more effort directed to goal-relevant activities than easy goals \citep{locke2002building}. Manifested in the findings from \cite{freiberg2021science}, firms with SBTs indeed make more investments in implementing carbon reduction projects. Further, the goal-setting theory suggests that the goal-performance relationship is the strongest when people are committed to the goal, especially for difficult goals \citep{locke2002building}. The SBTi publishes all firms' names with SBTs on its website and releases annual progress reports to evaluate these firms' performance against their SBTs. Once SBTs are set, firms are under public scrutiny, and failure to adhere to their targets may lead to detrimental consequences like damaged credibility and reputation. Therefore, firms adopting SBTs are typically more committed and motivated to strive for their targets. Finally, feedback is essential in the goal-setting theory to inform discrepancies between the goal and performance so that people can take corrective actions accordingly. The annual progress report released by the SBTi serves as a good feedback tool that firms can use to benchmark their progress and, hopefully, propose solutions to reduce performance discrepancies. Following the discussion, we posit that:

\begin{hypothesis}\label{hypo:carbonperf}
Adopting SBTs improves firms' carbon performance.
\end{hypothesis}

Although SBTs inform firms about how much carbon emissions to reduce, they do not provide guidance about how to achieve the goal. Reducing carbon emissions is a complex task that involves intra- and inter-organizational collaborations. Complex tasks can negatively affect goal effectiveness if not accompanied by higher-level skills and effective strategies \citep{wood1987task}, to the extent that the negative impact can offset the positive effect of goal difficulty on task performance. However, when effective strategies are used, the goal-performance relationship becomes stronger \citep{durham1997effects}. In the context of achieving SBTs, firms with a track record of superior environmental performance, i.e., the best-in-class, have an advantage because they are most likely equipped with the appropriate resources and strategies to address environmental issues. For instance, \cite{eccles2014impact} find that high sustainability firms -- i.e., the firms that voluntarily adopt sustainability policies by 1993 -- are more likely to have board members responsible for sustainability and top executive compensation incentives linked to sustainability metrics. Further, firms with ample experience in carbon reduction can quickly navigate through a variety of carbon abatement strategies and choose the most effective ones for them \citep{blanco2022classification}. In contrast, firms that do not belong to the best-in-class group may have to undergo extensive trial and error before they can find the best strategy. Following this line of reasoning, we hypothesize that:
\setcounter{hypothesis}{2} 
\begin{subhypothesis}\label{hypo:best}
\textit{The improvement of carbon performance due to adopting SBTs is more significant for best-in-class firms in terms of relative carbon performance.}
\end{subhypothesis}

Another stream of literature on operations strategy tells a different story. As firms must maintain business viability while pursuing environmental strategies, they tend to prefer carbon reduction initiatives with short payback periods \citep{blanco2020carbon}. However, firms may find it increasingly challenging to reduce carbon emissions as they move beyond the low-hanging fruit and start to face increasing costs and diminishing returns \citep{wu2011balancing}. The best-in-class firms, thus, may find themselves in this situation when adopting SBTs. In contrast, firms with little experience in carbon emissions reduction have a higher potential to reduce carbon emissions. This leads us to hypothesize that:
\begin{subhypothesis}\label{hypo:worst}
\textit{The improvement of carbon performance due to adopting SBTs is more significant for worst-in-class firms in terms of relative carbon performance.}
\end{subhypothesis}

Note that Hypothesis~\ref{hypo:best} and \ref{hypo:worst} are contrasting hypotheses supported by two mechanisms that can take effect simultaneously. Whether one mechanism dominates another or the effects of two mechanisms cancel each other out is an empirical question that demands a formal analysis.  
\subsection{Managerial Attention and Trade-offs in Sustainability}\label{subsec:tradeoffs}
The attention-based view of the firm proposes that what decision-makers accomplish depends on what issues they direct attention to \citep{ocasio1997towards}. The attention is further influenced by the environment decision-makers are situated in and may shift toward and away from certain issues and solutions based on the stimuli received from the environment. In the literature, \cite{tyre1994windows} find that the window of opportunity for technological adaptation closes when production pressure diverts the resources from problem-solving and continuous improvement of a newly implemented technology to daily production. \cite{dhanorkar2018promoting} stress that punitive tactics like inspections can redirect managerial attention and thus affect the efficacy of environmental improvement projects. 

In the context of the triple bottom line, numerous studies have examined the trade-offs between the economic and environmental bottom lines \citep{king2001lean,trumpp2017too,wagstaff2020tensions,awaysheh2020relation}. We argue that the attention-based view of the firm is helpful in understanding potential trade-offs between improving carbon performance and other non-carbon sustainability metrics. Global warming affects the population worldwide due to extreme weather caused by increasing temperatures, thus creating the urgency of having a global standard like SBTs to guide the progress of carbon emissions reduction. Other sustainability issues like water stress, electronic waste, and community conflicts are arguably more local problems that do not receive attention comparable to global warming. As a result, firms devoting resources to addressing carbon emissions problems may devote less attention on initiatives related to other non-carbon sustainability metrics. 

Another mechanism that may affect the performance of non-carbon sustainability metrics is the interdependence between reducing carbon emissions and using other resources. For example, the promotion of biofuels and carbon capture and storage by the Australian government is criticized due to the significant increase in water use \citep{pittock2013australian}. The transition from gasoline-based vehicles to electric vehicles, although favorable for reducing carbon emissions in the existence of relatively efficient electricity generation, creates a serious waste management challenge for end-of-life batteries \citep{harper2019recycling}. Research also finds that the implementation of the carbon price policy, if not carefully considered, can negatively affect food security and contribute to deforestation \citep{liu2019identifying}.

Adopting and achieving SBTs is a non-trivial task. It requires significant effort from firms to calculate current carbon emissions levels for their businesses and supply chains. Once targets are set, firms under public scrutiny must set up a budget along with a roadmap to achieve SBTs and report their progress on carbon emissions reduction periodically. Due to the ambitiousness of SBTs, significant effort is required to implement the carbon reduction plan. Therefore, we posit that firms with SBTs direct more attention to carbon performance. The increased attention can lead to more effort in carbon reduction projects and less effort in other non-carbon sustainability projects. As a result, the overall non-carbon sustainability performance can suffer, owing to both shifted attention and resources and to the potential negative impact carbon abatement projects can have on other sustainability metrics, as discussed above. We formalize this hypothesis as:
\begin{hypothesis}\label{hypo:noncarbonperf}
Adopting SBTs is associated with a decrease in non-carbon-related sustainability performance.
\end{hypothesis}

\section{Data and Methodology}\label{sec:method}
\subsection{Data}\label{subsec:data}
We combine two data sets for our empirical analysis. We use the information provided by the SBTi to obtain the data on firms' SBT approval dates. A firm must submit a carbon emissions reduction target that covers one or more of the following scopes to the SBTi for evaluation. Scope 1 emissions are direct greenhouse gas emissions from sources that are owned and controlled by the firm. Scope 2 emissions are indirect emissions associated with purchased electricity, steam, or heat that occur outside the firm's boundary. Scope 3 emissions are those that occur due to activities up and down the firm's supply chain, such as emissions from the production of purchased materials and product use by customers. Once approved, the target will be announced as science-based, and the firm's name will appear on the SBTi's website as Companies Taking Action (https://sciencebasedtargets.org/companies-taking-action). In addition to the science-based standard, the SBTi also documents the temperature alignment of an SBT (i.e., 2 degrees Celsius, well-below 2 degrees Celsius, or 1.5 degrees Celsius) and whether the firm has a long-term or net-zero target. 

Firms' financial and Environmental, Social, and Governance (ESG) data are acquired from Refinitiv Workspace (formerly Thomson Reuters). Refinitiv Workspace is a terminal that provides broad coverage of financial data, news, and analytics tools. It offers one of the most comprehensive ESG databases. The firm catalog represents over 85\% of the global market cap across 76 countries. The ESG data are collected from multiple sources like firms' annual financial and Corporate Social Responsibility (CSR) reports, Non-Government Organizations (NGO) websites, stock exchange filings, and news sources. Data quality is ensured through systematic pre- and post-production error checks and independent audits \citep{refinitiv2022esg}.

For our sample, we first identify public firms that disclose their scope 1 and 2 $CO_2$ equivalent emissions in their latest fiscal year. In this study, we do not consider scope 3 emissions for two reasons. First, the SBTi's requirement for setting a scope 3 emissions target is less rigid compared to that for scope 1 and 2 emissions. In particular, based on the SBTi Corporate Manual \citep{sbti2021corporate}, firms must set SBTs that cover at least two-thirds of scope 3 emissions only if scope 3 emissions exceed 40\% of their total emissions. Firms are also allowed to use methods other than the two SBTi-recommended ones (i.e., the ACA and the SDA) for scope 3 emissions target-setting. These differences in target-setting standards can confound the results of our analysis. Second, the disclosure of scope 3 emissions is less common than that of scope 1 and 2 emissions. Even if disclosed, the scope 3 emissions tend to be less accurate due to the complex nature of a supply chain \citep{blanco2016state}. An increase in scope 3 emissions, for example, could be because the firm has identified new sources of carbon emissions in its supply chain or simply due to a change of estimation method \citep{bjorn2022can}. We also observe that scope 3 emissions tend to be less consistent, if not completely missing, for a significant portion of the firms in our sample. Including scope 3 emissions may introduce selection bias and measurement error. Thus, we exclude scope 3 emissions in this study. The initial sample includes 4,163 unique firms (excluding the financial sector). 

We use ISIN as the identifier to identify whether a firm in our initial data set has an established SBT. As the SBTi does not report ISIN for all listed firms, we use additional information like firms' names, sectors, headquarter locations, etc., to identify firms with SBTs.

\subsection{Model Specification}\label{subsec:model}
Firms adopt SBTs at different points in time and remain committed at all times afterwards. The staggered SBT adoption allows us to use a regression-based difference-in-differences (DiD) approach for causal inference. The model is specified as follows:
\begin{align*}
    Y_{it}=\alpha_i+\delta_t+\beta(Treated_i\times Post_t) + \gamma C_{it}+\epsilon_{it},
\end{align*}
where $Y_{it}$ is the dependent variable for firm $i$ at time $t$, $\alpha_i$ is the firm fixed effect, and $\delta_t$ is the year fixed effect. $Treated_i$ is a binary variable that indicates the treatment groups, and it equals one in all firm-year observations if the firm adopts SBTs. $Post_t$ is an indicator that equals one if period t is at least one year after the adoption of SBTs. Finally, time-varying control variables \textbf{$C_{it}$} are included to control for factors that change over time, and $\epsilon_{it}$ is the error term. This model compares the change in carbon emissions for firms before and after adopting SBTs (the treatment group) to the change for firms that have not adopted SBTs yet (the control group). $\beta$ is the coefficient of interest that estimates the \textit{average treatment effect on the treated} (ATT): the effect of adopting SBTs on the dependent variable.

\subsubsection{Propensity Score Matching}\label{subsec:psm}
The DiD approach resolves the identification challenge of non-observable counterfactuals -- the potential outcomes of the control group if treated and the potential outcomes of the treatment group if not treated. In particular, it uses the outcomes of the control group as the counterfactual outcomes of the treatment group. This approach rests on the parallel trend assumption: the trend of the observed outcomes of the control group resembles that of the treatment group, had firms in the treatment group not been treated. 

The parallel trend assumption may be violated if firms in the control group fundamentally differ from those in the treatment group. In this study, firms with SBTs may have selected themselves into this decision and thus demonstrate distinct characteristics from those without SBTs. This selection bias can confound the effect of having SBTs on carbon emissions. We use propensity score matching to address this problem. The propensity score matching technique matches firms in the treatment and control groups with a similar propensity to be treated based on several observable firm characteristics (i.e., covariates). It reduces the heterogeneity between the control and treatment groups and helps address endogeneity concerns stemming from the selection bias. The matched sample allows us to make stronger claims about the causal influence of SBTs.

To identify relevant covariates to be used in the matching process, we perform a logit regression to determine the drivers of adopting SBTs (details reported in \S \ref{subsec:prelimresults}). We find that firms' total revenue, number of employees, environmental pillar scores, profit margin, and combined scope 1 and 2 carbon emissions are significantly associated with their probability of adopting SBTs (the results of this analysis are in Table~\ref{tab:driverSBT}).  

\begin{figure}[h!]\caption{Data cleaning procedures}\label{fig:data}
\begin{center}
\includegraphics[scale=0.5]{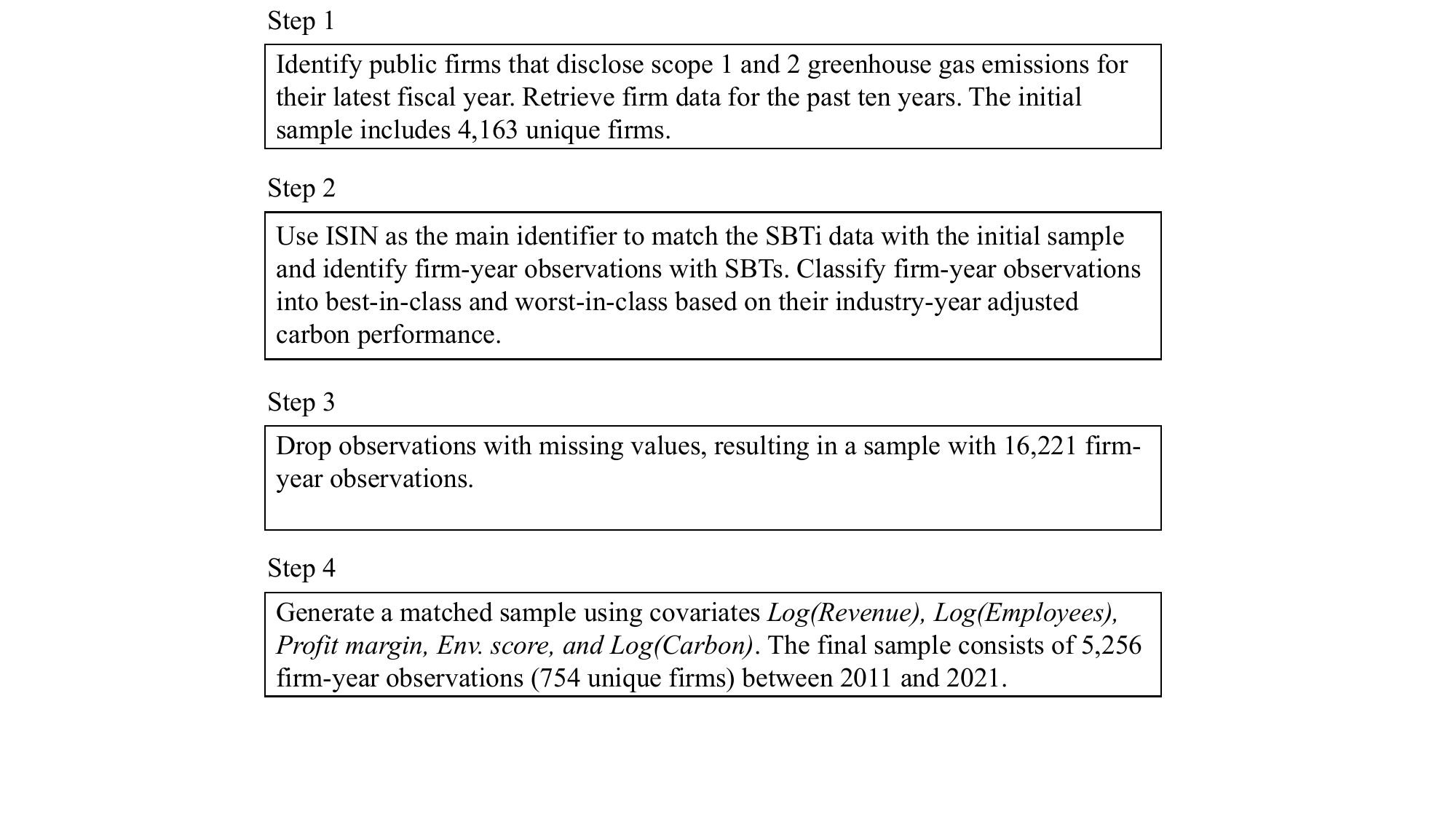}
\end{center}
\vspace{-0.5in}
\end{figure}

One issue with the staggered DiD is the existence of problematic $2\times 2$ comparisons in calculating the treatment effect. For example, \cite{goodman2021difference} find that one component in the two-way fixed effect DiD estimator uses the post-treatment periods of the earlier-treated units as controls to which the later-treated units are compared when they receive treatment; it confounds the true treatment effect. We use the stacked regression estimator suggested by \cite{baker2022much} to alleviate this concern. We first group the treatment units by the year they are treated. Second, using the nearest neighbor algorithm, we match the value of the covariates one year prior to the treatment year to identify the control units (i.e., the nearest neighbor based on the propensity to be treated) for each treatment unit. We keep the observations between seven years before and five years after treatment. The process is repeated for each year between 2016, the earliest year firms' SBTs can be approved, and 2021. The match is conducted with replacement, and we allow the same control unit to be matched to multiple treatment units. Finally, we stack all treatment and control units together to form the new sample. About two-thirds of the observations were dropped in the process either due to missing data or because they were not matched to any other units. Figure~\ref{fig:data} shows the step-by-step data clearing procedures. Our final sample consists of 754 unique firms and 5,256 firm-year observations across 21 industry groups (by four-digit GICS codes) and 43 countries (see Table~\ref{tab:desriptive_industry} and \ref{tab:desriptive_country} for the number of firms by industry and country). The number of companies adopting SBTs is 377, comparable to existing empirical studies about SBT adoptions and effects \citep{giesekam2021science,freiberg2021science}. Overall, our sample demonstrates a diverse set of firms representative of the population.

Table~\ref{tab:matching} shows the matching results comparing the mean of covariates before and after matching. The slight difference in mean values of the treatment groups between the full and matched samples is due to the matching process explained above. Observe that the matched sample significantly reduced the differences between the control and the treatment units regarding the observed covariates. Note that although the difference in carbon emissions between the treatment and control groups is still significant after matching, we will control for this difference using firm and year fixed effects in the main analysis. Thus, the matching results increase our confidence in using the control group as the counterfactual to the treatment group. 

\input{matching.tex}

\subsection{Measures}\label{subsec:measures}
\subsubsection{Dependent Variables}\label{subsec:dv}
The primary dependent variable, \textit{Log(Carbon)}, is the log-transformed value of combined scope 1 and scope 2 emissions. This variable measures the absolute carbon performance of a firm, as opposed to the relative intensity measure (e.g., carbon emissions generated per unit of sales). This is consistent with the premise of SBTs to reduce absolute carbon emissions regardless of economic growth. This measure is frequently used in the literature to assess the effectiveness of various carbon emissions reduction-related practices \citep{dahlmann2019managing,fu2020take,park2022vessel}. In \S \ref{subsec:robustness}, we perform the same analysis with several alternative dependent variables with respect to carbon performance.

To test Hypothesis~\ref{hypo:noncarbonperf} regarding SBTs' impact on non-carbon sustainability performance, we use Refinitiv's ESG scores as dependent variables. Refinitiv publishes ESG scores based on firms' relative performance of ESG factors with the firm's sector \citep{refinitiv2022esg}. It uses over 630 firm-level measures to arrive at the performance scores under three pillars: environmental, social, and governance. The evaluation of ESG performance accounts for the importance of various ESG measures across industries by applying industry-specific weights, thus reducing industry and transparency biases. 

In particular, we use the environmental and social pillar scores in our analysis to represent the overall sustainability performance of a firm. The environmental pillar score evaluates firms' performance under three categories: emissions, resource use, and innovation. The data points used for evaluation include carbon emissions, water withdrawal and discharge, and waste reduction, among others. As carbon performance will also affect the environmental pillar score, the premise of using this score is that adopting SBTs will have a non-negative impact on firms' carbon performance. Thus, if the environmental pillar score becomes lower after adopting SBTs, it is highly likely that non-carbon sustainability performance has decreased. 

The social pillar score represents another non-carbon sustainability performance. It is evaluated using four categories: workforce, human rights, community, and product responsibility. Examples of measures for calculating the score include health and safety policies, human rights policies, diversity and equal opportunity policies, community donations, ISO 9000, and quality management programs. The score represents firms' overall performance in social responsibility. 
\subsubsection{Independent Variables}\label{subsec:iv}
Following the basic DiD model \citep{angrist2009mostly}, we divide firms into two groups: a treatment group for firms with SBTs, and a control group for firms without SBTs. We capture firms' status using two binary variables, $Treated$ and $Post$. $Treated$ equals one when a firm belongs to the treatment group and zero otherwise. For firms in the treatment group, $Post$ equals one for the periods where a firm has already adopted an SBT and zero otherwise. We match a control firm to each treated firm and assign the same value to $Post$ for control firms as their treated counterparts. 

To identify the best-in-class and worst-in-class firms in terms of carbon performance, we follow existing literature \citep{voss1997benchmarking,awaysheh2020relation} and use a 10\% cut-off point (alternative cut-off points are examined in \S \ref{subsec:robustness}). Considering the impact of time and industry context on firms' carbon performance, we categorize a firm as best-in-class (worst-in-class) if its combined scope 1 and 2 emissions level is lower (higher) than the ten (ninety) percentile level relative to its industry peers (based on the four-digit GICS codes) in a given year. Using a 10\% cut-off point ensures that firms that are identified as best-in-class or worst-in-class are significantly different from others in terms of carbon performance, thus in line with our purpose to compare their performance with their peers after adopting SBTs. 
\subsubsection{Control Variables}\label{subsec:cv}
We control for variables that may confound the relationship between SBT adoption and a firm's carbon performance. We first identify factors that may affect the dependent variables. Firm size has been shown to affect corporate carbon performance \citep{hofer2012competitive}. We use the log-transformed number of employees and total revenue to control for firm size. In addition, a firm's financial performance and operational efficiency can be associated with its carbon performance \citep{jacobs2016operational,awaysheh2020relation,bellamy2020administrative}. Therefore, we use several metrics, including sales growth, profit margin, and return on assets to control for firms' financial performance, and inventory turnover to control for operational efficiency. Further, as firms' overall environmental performance can affect their carbon emissions, we include Refinitiv's environmental pillar score as a control variable. Based on the analysis of drivers of SBT adoption (see \S \ref{subsec:prelimresults} for details), we retain total revenue, number of employees, profit margin, and environmental pillar score as controls in our analysis, as these variables are also significantly associated with firms' SBT adoption decisions. 

Finally, we include firm fixed effects to control for time-invariant factors that may affect one's carbon performance. We also include year fixed effects to control for potential influence from temporal factors (e.g., overall economic conditions). Table~\ref{tab:correlation} shows the descriptive statistics and correlation matrix after matching. 

\input{correlationresults.tex}

We use ordinary least squares (OLS) on the matched sample to obtain results. We cluster standard errors at the firm level to account for heteroskedasticity and autocorrelation. As a diagnostic test, we calculate the variance inflation factors (VIFs) for independent and control variables. We find that the average VIF is 1.54, and none of the individual VIFs exceeds 3, indicating that multicollinearity is not a serious concern.

\section{Results}\label{sec:result} 
This section briefly discusses the drivers of adopting SBTs and then reports the main results using the matched sample to answer the research questions. 
\subsection{Drivers for SBT Adoption}\label{subsec:prelimresults}
As described in \S \ref{subsec:psm}, we estimate a logit model to identify the firm characteristics that drive their decision to adopt SBTs. Previous literature has found that firms are more likely to adopt carbon targets when they are larger and have better historical ESG and financial performance \citep{giesekam2021science,freiberg2021science}. In our model, we use \textit{Log(Revenue)}, \textit{Log(Employees)} and \textit{Sales growth} to represent firms' size and growth potential, \textit{Env. score} and \textit{Log(Carbon)} for firms' overall environmental and carbon performance, and \textit{Profit margin} and \textit{ROA} for firms' financial performance. In addition, we add \textit{Inventory turnover} as a common proxy for operational performance. We extract the data one year before firms' SBT adoption and pool the observations of firms adopting SBTs at different times in the same sample. We further control year fixed effects to estimate our model. Results of this analysis are shown in Table~\ref{tab:driverSBT}. 

\input{logit.tex}

Consistent with the literature, we find that larger firms, indicated by higher revenue and more employees, are more likely to adopt SBTs. Further, higher profit margins and environmental scores are also associated with a higher probability of adopting SBTs. Interestingly, based on our analysis, firms with higher levels of carbon emissions are less likely to engage in SBTs, echoing the concern that firms with SBTs may account for only a small share of global emissions \citep{trexler2015science}. 

Our results show that there is indeed a selection bias regarding SBT adoption that, if not controlled, will confound the relationship between SBT adoption and future carbon performance. Thus, in our main analysis concerning the effectiveness of SBTs, we focus on the more balanced matched sample and control for significant predictors (i.e., \textit{Log(Revenue)}, \textit{Log(Employees)}, \textit{Profit margin}, \textit{Env. score} and \textit{Log(Carbon)}) to mitigate endogeneity concerns.
\subsection{Main Results}\label{subsec:mainresults}
This section discusses the models we used for hypothesis testing and the corresponding results obtained from our analysis. We first examine the effect of SBTs on carbon emissions reduction, with the main results reported in Table~\ref{tab:resultscarbon}. To test Hypothesis 1, we start with the basic two-way fixed effect DiD model \citep[Equation~\ref{eqn:h1nc},][]{angrist2009mostly}. 
\begin{align}\label{eqn:h1nc}
    Log(Carbon)_{it}=&\alpha_i Firm_{i}+\delta_t Year_{t}+\beta(Treated_i\times Post_t)+\epsilon_{it}.
\end{align}
Here, $Firm_i$ is the firm dummy variable, $Year_t$ is the year dummy variable, $Treated_i$ is an indicator of treated firms that adopted SBTs at some point, and $Post_t$ denotes the post-treatment periods for both control and treatment groups. The coefficient of interest is $\beta$, and recall that following Hypothesis 1, we expect to see a negative $\beta$. Observe from Model (1) in Table~\ref{tab:resultscarbon} that the coefficient of the interaction term $Treated \times Post$ is negative and significant ($\beta=-0.103$, $p<0.05$), indicating that adopting SBTs leads to reduced carbon emissions.

Equation~\ref{eqn:h1nc} controls for time-invariant effects, such as individual firms' characteristics, but does not address omitted variable bias from time-varying variables. Equation~\ref{eqn:h1fe} adds covariates identified in \S \ref{subsec:prelimresults} to control for such bias \citep{angrist2009mostly}. We do not include $Log(Carbon)$ in the control variables as (1) it is the dependent variable of the model, and (2) the DiD model has already accounted for the changes in carbon emission across firms and over time. Observe from Model (2) in Table~\ref{tab:resultscarbon} that the coefficient of interest remains negative, the effect size decreases, and it is no longer significant ($\beta=-0.089$, $p>0.05$). 
\begin{align}\label{eqn:h1fe}
        Log(Carbon)_{it}=&\alpha_i Firm_{i}+\delta_t Year_{t}+\beta(Treated_i\times Post_t) + \gamma Control_{it}+\epsilon_{it}.
\end{align}

\begin{figure}[h!]
\caption{Dynamic DiD coefficients with 95\% confidence intervals}\label{fig:dyncoef}
\begin{center}
\includegraphics[scale=0.75]{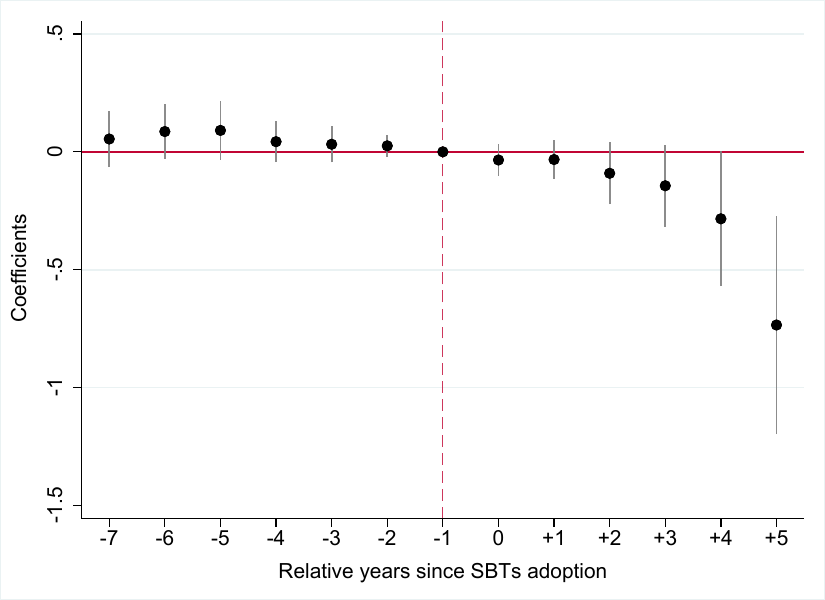}
\end{center}
\end{figure}

What we find in Model (2) implies that omitted variable bias could be an issue that leads to a biased coefficient in Model (1), calling on more analysis for model validation. We use the relative time model (Equation~    \ref{eqn:h1dy}) to establish causality by confirming the parallel assumption \citep{granger1969investigating}.
\begin{align} \label{eqn:h1dy}
    Log(Carbon)_{it}=&\alpha_i Firm_{i}+\delta_t Year_{t}+\beta_k(Treated_i\times\sum_{k=-7}^{-2} Pre_t^k) \nonumber\\
    &+\beta_l(Treated_i\times\sum_{l=0}^{5} Post_t^l) + \gamma Control_{it}+\epsilon_{it}. 
\end{align}
Equation~\ref{eqn:h1dy} is the DiD model that estimates the dynamic ATTs by interacting \textit{Treated} with relative time indicators (i.e., time periods leading to and following treatment). In Equation~\ref{eqn:h1dy}, $\sum_{k=-7}^{-2} Pre_t^k$ is a set of binary variables that specify pre-treatment time periods, while $\sum_{l=0}^{5} Pre_t^l$ is a set of binary variables for post-treatment time periods. We consider up to seven years before and up to five years after treatment and omit the year prior to treatment as the base year. Observe from Model (3) in Table~\ref{tab:resultscarbon} that the coefficients for pre-treatment time periods are insignificant (thus collapsed into one row), supporting the parallel trend assumption. Further, all post-treatment coefficients are negative, and the effect size increases as firms move forward with their SBTs. However, only the coefficient for \textit{Post-treatment(t+5)} is statistically significant ($\beta=-0.735$, $p<0.01$). Figure~\ref{fig:dyncoef} provides a graphical illustration of the coefficients with their 95\% confidence intervals for the relative ATTs. The figure confirms the parallel trend preceding SBTs adoption and the decreasing trend following SBTs adoption. The number of observations decreases over time, and only six firms adopting SBTs in 2016 have data five years after treatment (i.e., \textit{t+5}). The small fraction of firms achieving significant carbon emissions reduction does not result in an overall significant ATT, as observed in Model (2). Therefore, we conclude that Hypothesis 1 is partially supported.
\vspace{-0.2in}
\begin{align}\footnotesize
    Log(Carbon)_{it}=&\alpha_i Firm_{i}+\delta_t Year_{t}+\beta(Treated_i\times Post_t\times Worst\,in\,class_{it}) \nonumber\\
    &+ \gamma Control_{it}+\epsilon_{it}, \label{eqn:h2worst}\\
    Log(Carbon)_{it}=&\alpha_i Firm_{i}+\delta_t Year_{t}+\beta(Treated_i\times Post_t\times Best\,in\,class_{it}) \nonumber\\
    &+ \gamma Control_{it}+\epsilon_{it}. \label{eqn:h2best}
\end{align}
Equation~\ref{eqn:h2worst} and \ref{eqn:h2best} are constructed to test Hypothesis~2A and 2B. We use the three-way interaction terms $Treated_i\times Post_t\times Worst\,in\,class_{it}$ and $Treated_i\times Post_t\times Best\,in\,class_{it}$ to estimate the treatment effects of the best- and worst-in-class firms relative to the rest of the sample. Models (4) and (5) in Table~\ref{tab:resultscarbon} report the results of testing Hypothesis 2A and 2B. The coefficients of the interaction terms $Treated \times Post \times Worst\,in\,class$ and $Treated \times Post \times Best\,in\,class$ are both negative but insignificant ($\beta=-0.175$, $p>0.05$; $\beta=-0.259$, $p>0.05$;). These results indicate that firms' relative carbon performance compared to their industry peers does not significantly influence their carbon emissions reduction after adopting SBTs. Therefore, Hypothesis 2A and 2B are not supported.

\input{h1h2results.tex}

\input{h3results.tex}

Table~\ref{tab:resultsnoncarbon} shows the effects of SBTs on overall environmental and social performance (Hypothesis 3). We use the two-way fixed effects model with controls (i.e., Equation~\ref{eqn:h1fe} with different dependent variables; equations are not reported to avoid duplication) to estimate the ATTs. Recall that Hypothesis 3 suggests that adopting SBTs can negatively affect non-carbon-related sustainability performance. Thus, we expect to find a negative relationship between the $Treated \times Post$ interaction term and the dependent variables. In both models, we find that the coefficients of the interaction term are negative but insignificant ($\beta=-1.065$, $p>0.05$ for \textit{Env. score}; $\beta=-1.146$, $p>0.05$ for \textit{Social score};). It indicates that adopting SBTs does not significantly impact firms' overall environmental or social performance. Therefore, Hypothesis 3 is not supported.

\subsection{Robustness Check}\label{subsec:robustness}
We report the results of several robustness tests in this subsection. We first perform the same analysis using four alternative dependent variables for carbon performance. In particular, we use $\frac{Carbon_{i}-\mu(Carbon)_{jt} }{\sigma(Carbon)_{it}}$ to indicate the standardized value of carbon emissions by industry \textit{j} and year \textit{t} and $Log(\frac{Carbon_{it}}{Revenue_{it}})$ to compute the log-transformed carbon intensity measure that accounts for firms' annual sales volume. These two variables measure the relative carbon performance, as opposed to the absolute carbon performance we used for the main analysis, of a firm.  We further use $100\times\frac{Carbon_{i,t+1}-Carbon_{it}}{Carbon_{it}}$ and $100\times\frac{Carbon_{it}-Carbon_{i,t-1}}{Carbon_{i,t-1}}$ to represent the forward-looking and backward-looking annual percentage change in carbon emissions at the firm-year level. All these measures are commonly used in both industry and academia \citep{dahlmann2019managing,fu2020take}. Despite the different meanings of these measures, the main results are consistent with those in the original analysis: adopting SBTs does not have a significant impact on carbon performance. For brevity, we report the static and dynamic DiD model results without interactions with best- and worst-in-class firms in Tables~\ref{tab:resultscarbonstddv} and \ref{tab:resultscarbonratiodv}.

\input{h1h2resultscem}

In addition to propensity score matching, we performed a non-parametric matching algorithm -- the coarsened exact matching \citep[CEM;][]{blackwell2009cem}. CEM coarsens continuous variables into bins that contain subsets of each covariate's value used for matching. It then creates one stratum for each combination of bins and places every observation in a stratum. The observations in the same stratum will be matched. We use the same covariates (\textit{Profit margin, Log(Revenue), Log(Employees), Env. score, Log(Carbon)}) as matching covariates. As not all treated units are guaranteed to have a match, the number of treated firms after matching reduces to 172, less than half of the treated firms in the main analysis. The results obtained from a CEM-matched sample are reported in Table~\ref{tab:resultscarboncem} (we omitted the results for testing Hypothesis 3 as they are consistent with the main analysis). Interestingly, in the basic DiD model with controls (Column (2)), we find that the coefficient of \textit{Treated$\times$Post} is negative and significant ($\beta=-0.098$, $p<0.05$;), indicating that adopting SBTs results in lower carbon emissions. A further observation from the dynamic DiD model (Column (3)) shows that only firms in their fourth year after adopting SBTs (\textit{Treated$\times$Post-treatment(t+4)}) appear to make a difference in carbon emissions ($\beta=-0.309$, $p<0.05$). That is a small fraction of treated firms (9 out of 172) in our sample. Finally, in Column (5), we find that the coefficient of \textit{Treated$\times$Post$\times$Best in class} is negative and significant ($\beta=-0.362$, $p<0.05$), providing initial support to Hypothesis 2A that best-in-class firms improve carbon performance more significantly after adopting SBTs compared to the rest of firms.

Given the results using CEM, we further explore if changing the cut-off point of classifying best- and worst-in-class firms would impact our results in the PSM-based model. In addition to the original 10-percent cut-off point used in the base model, we tested Hypothesis 2 using alternative cut-off points at 2.5, 5, 15, 20, and 25 percent of industrial peer-adjusted carbon performance, respectively. We found that for any percentage less than or equal to 15, being classified as best- or worst-in-class does not affect firms' carbon performance after adopting SBTs. This is consistent with our observation in the main analysis. However, when we expand the best-in-class group to the top 20 and 25 percent, we find that best-in-class firms have better carbon performance after adopting SBTs than the rest of the sample. As shown in Table~\ref{tab:resultscarbonaltperc}, the coefficients for the three-way interaction terms in Column (2) and (4) are negative and significant ($\beta=-0.246$, $p<0.05$; $\beta=-0.195$, $p<0.05$).

\input{h1h2resultsaltperc}

To further reduce the concern that the relationship between SBT adoption and carbon emissions reduction is driven by some unobservable time-variant characteristics, we conducted a placebo test \citep{dhanorkar2019environmental,song2020value}. In this test, we hold the number of treatment units constant and randomly assign pseudo-treatment to firms for each treatment year based on the uniform distribution. We re-estimated the DiD model and did not find a significant relationship between our main variables of interest (results omitted for brevity). This increases our confidence in the causal impact of SBTs on carbon emissions.

\section{Discussion}\label{sec:discussion}
The debate on the effectiveness of SBTs has caught the attention of both industry and academia since the Paris Agreement. In this study, through extensive analyses using a sample of international firms across various industries, we found mixed evidence regarding the effectiveness of SBTs. We demonstrated that although a small number of firms took four to five years to improve their carbon performance after adopting SBTs, this did not always translate to a more general treatment effect for all treated firms. Further, the carbon emissions reduction effect is more prominent for the best-in-class firms. We also found that adopting SBTs does not have a significant impact on firms' overall environmental and social responsibility performance. 

Theoretically, we contribute to the literature by extending the goal-setting theory to explain firms' ESG performance. Our results reveal that having a specific and challenging carbon target does not always enhance firms' carbon performance. This contrasts the existing findings that firms with SBTs invest more in carbon initiatives \citep{freiberg2021science} and report more aggressive carbon emissions reduction \citep{SBTi2022progress}. There are two potential explanations. On the one hand, our results show that a few early adopters of SBTs have achieved significant carbon emissions reduction four to five years after their adoptions. As the SBTi has been certifying SBTs only since 2016, the early adopters only account for a small fraction of SBT firms. Thus, the effect of SBTs could be more salient in the following years when the number of firms sticking to their carbon commitment increases and hits a critical mass. On the other hand, as SBTs become more and more influential over time, many late adopters may choose, or be forced, to adopt SBTs for marketing purposes: having SBTs sends a positive signal about their ESG commitment. Thus, they may not be as committed as the early adopters to pursuing these targets. SBTs then become symbolic and do not elicit more attention or effort for goal achievement as predicted by the goal-setting theory. The findings in this study provide a middle ground in the debate on the effectiveness of SBTs, and it is probably too early to tell whether SBTs can materialize beyond a specific group of firms. 

Further, we contribute to the sustainable operations literature by examining potential trade-offs between sustainability metrics. Although both the literature and industrial practices suggest that managerial attention shifted to one project may reduce resources available for other projects, the existing evidence is more focused on trade-offs between specific sustainability programs \citep{dhanorkar2019environmental} or the United Nations Sustainable Development Goals at the macro level \citep{parkinson2019balancing}. In contrast, we examined the trade-offs at the firm level and did not find evidence that adopting SBTs would negatively affect firms' overall environmental or social performance. Our interviews with senior managers at Flex Ltd. revealed one possible explanation for this phenomenon. One of the senior managers commented that Flex had well-established sustainability policies and did not believe they would be easily affected by a project like SBT \citep{flex2022}. It seems that although these trade-offs might exist locally, the magnitude is not significant enough to affect firms' environmental and social sustainability performance at an aggregate level.

Our work also yields several managerial and political implications. First, we found that it would take years before firms could see decreasing carbon emissions due to SBTs adoption. As firms are often resource-constrained and results-driven, they may terminate the SBT project prematurely due to the lack of progress in the first year and miss the real benefits in the following years. Therefore, it is crucial for SBT proponents to make managers aware of this delay between SBT implementation and benefits realization.    

Second, our findings reveal that firms with superior carbon performance are more likely to adopt SBTs, and having SBTs does provide additional benefits in terms of carbon emissions reduction for the best-in-class firms. It reconciles the doubt that those best-in-class firms adopting SBTs are already on the right track to reduce carbon emissions and will not make any additional effort or progress after adoption \citep{trexler2015science}. In fact, even if some of these firms already have internal carbon targets, substituting these targets with SBTs can create new momentum due to increased target difficulty \citep{freiberg2021science} and exposure to public scrutiny. Thus, best-in-class firms are encouraged to take advantage of SBTs, if they have not already done so, to further enhance their carbon performance.

Further, our results raise concerns regarding the worst-in-class firms, as (1) they are less likely to adopt SBTs, and (2) even if they do, their carbon performance may not improve after adoption. One possible explanation for this phenomenon is that SBTs are too challenging for these firms to achieve within the specified time frame. According to the goal-setting theory, overly aggressive goals can lead to poorer performance than no goals if people lack the knowledge to achieve these goals \citep{latham2006enhancing}. In the context of this study, firms performing poorly in the past may lack the organizational structure and knowhow to reduce carbon emissions, resulting in little progress after adopting SBTs. This is a discouraging finding, but it conveys an important message that having SBTs itself may not be sufficient for some firms. Firms need to align operations goals with strategic goals and actively implement carbon emissions reduction programs in order to make progress. Research suggests that a specific learning goal can be more effective than performance goals under these circumstances. To this end, the SBTi can hold panels and invite best-in-class firms to share their best practices in reducing carbon emissions. These activities can facilitate learning in worst-in-class firms, ideally before they commit to an SBT. 

Lastly, we did not find evidence that having SBTs would compromise firms' overall environmental and social sustainability performance. This is good news for firms considering adopting SBTs. However, we do not claim that trade-offs among SBTs and other projects do not exist. We advise firms to be cautious after adopting SBTs and carefully monitor the progress of other non-carbon projects implemented at the same time as SBTs.

\section{Conclusion}\label{sec:conclusion}
Building on the goal-setting theory, this study empirically examines the effectiveness of SBTs, the moderating role of firms' relative carbon performance, and potential trade-offs between SBTs and other non-carbon sustainability metrics. We conclude by discussing the limitations of this study and potential future research directions.

Our work shows that the effect of SBTs becomes prominent after four years, and only a limited number of early-adopting firms in our sample have had SBTs for more than four years. A natural question following this finding is how long this effect will last, as SBTs require a long-term commitment from firms, and whether we will observe a similar effect for the late adopters. As more data become available in the future, and as firms get closer to the end period of their SBTs, it is worth repeating the analysis with the dynamic DiD model and further exploring whether firms are on track to achieving their SBTs. 

While our research examined the effectiveness of SBTs in reducing carbon emissions, the question of how firms with distinct carbon reduction strategies may perform differently remains unanswered. In practice, firms tend to prioritize carbon projects with short payback periods \citep{blanco2020carbon}, and research has shown that these projects do not yield the greatest carbon emissions reduction \citep{blanco2022classification}. Investing in renewable energy, for example, has a significant impact on reducing carbon emissions. Still, realizing the benefits takes a long time and may not be economically viable for many firms in the short term. How should firms choose from a variety of carbon projects that best suit their needs? Are there trade-offs and synergies among these projects? Is there a sequence of projects that maximizes firms' combined economic and environmental gains? Future research can answer these questions by acquiring and investigating project-level data.

Despite anecdotal evidence suggesting potential trade-offs among different sustainability metrics, we did not find supporting evidence in our analysis. This is possibly due to the lack of granular data reporting the performance of various programs. Future research can look for sustainability programs with salient conflicts of interest or resource competition with carbon emissions reduction and examine the impact of adopting SBTs on the outcomes of these programs.


\bibliographystyle{pomsref}

 \let\oldbibliography\thebibliography
 \renewcommand{\thebibliography}[1]{%
    \oldbibliography{#1}%
    \baselineskip14pt 
    \setlength{\itemsep}{10pt}
 }
\bibliography{ref1}



\ECSwitch 

\ECHead{E-Companion for The Impact of Carbon Targets on Firms' Carbon Performance}

\input{descriptive_industry}
\clearpage

\input{descriptive_country}
\clearpage

\input{h1h2results_stddv} 
\clearpage

\input{h1h2results_ratiodv} 
\clearpage

\input{matching}
\clearpage

\end{document}

%% file: matching.tex
\begin{table}[!h]\centering\footnotesize
\def\sym#1{\ifmmode^{#1}\else\(^{#1}\)\fi}
\caption{Matching Results \label{tab:matching}}
\sisetup{
            detect-all,
            table-number-alignment = center,
            table-figures-integer = 2,
            table-figures-decimal = 1,
            table-space-text-post = {\superscript{*}},
}
\begin{tabular}{l*{2}{|SSS}}
\toprule
                &\multicolumn{3}{c|}{Full Sample}          &\multicolumn{3}{c}{Matched Sample}       \\
                & {Treatment}&   {Control}&      {Diff}         & {Treatment}&   {Control}&      {Diff}         \\
\midrule
Log(Revenue)    &    22.959&    21.954&    -1.004\sym{***}&    22.958&    23.150&     0.192\sym{***}\\
Log(Employees)  &    12.097&    10.855&    -1.242\sym{***}&    12.091&    12.121&     0.030         \\
Env. score      &    71.917&    57.402&   -14.515\sym{***}&    72.311&    72.435&     0.123         \\
Profit margin   &    40.100&    34.658&    -5.442\sym{***}&    39.534&    38.788&    -0.747         \\
Log(Carbon)     &    13.023&    12.625&    -0.399\sym{***}&    13.048&    13.351&     0.303\sym{***}\\
\midrule
Observations    &     16221&          &                   &      5256&          &                   \\
\bottomrule
\multicolumn{7}{l}{\footnotesize * $ p<.05 $, ** $ p<.01 $, *** $ p<.001 $}\\
\multicolumn{7}{l}{\footnotesize Matching was performed using nearest one neighbor based on the propensity to be treated.}\\
\end{tabular}
\end{table}

%% file: correlationresults.tex
\begin{table}[htbp]\centering\footnotesize
\def\sym#1{\ifmmode^{#1}\else\(^{#1}\)\fi}
\caption{Descriptive Statistics and Correlation Matrix \label{tab:correlation}}
\begin{tabular}{l*{1}{
S[table-format=2.2]
S[table-format=2.2]
S[table-format=2.2]
S[table-format=2.2]
S[table-format=-1.2]
S[table-format=-1.2]
S[table-format=-1.2]
S[table-format=-1.2]
S[table-format=-1.2]
S[table-format=-1.2]
S[table-format=-1.2]
S[table-format=-1.2]
S[table-format=-1.2]
}}
\hline\hline
          &\multicolumn{13}{c}{}                                                                                                                      \\
          &     {Mean}&       {SD}&      {Min}&      {Max}&      {(1)}&      {(2)}&      {(3)}&      {(4)}&      {(5)}&      {(6)}&      {(7)}&      {(8)}&      {(9)}\\
\hline
(1) Log(Carbon)&    13.19&     2.17&     3.37&    18.97&         &         &         &         &         &         &         &         &                  \\
(2) Env. score&    72.37&    16.37&     6.38&    99.22&     0.23&         &         &         &         &         &         &         &                  \\
(3) Social score&    71.80&    16.81&    12.32&    99.56&     0.20&     0.54&         &         &         &         &         &         &                 \\
(4) Treated&     0.53&     0.50&     0.00&     1.00&    -0.07&    -0.00&     0.08&         &         &         &         &         &                  \\
(5) Post  &     0.28&     0.45&     0.00&     1.00&    -0.06&     0.14&     0.15&     0.01&         &         &         &         &                  \\
(6) Best in class&     0.09&     0.28&     0.00&     1.00&    -0.33&    -0.04&    -0.09&    -0.01&     0.01&         &         &         &                  \\
(7) Worse in class&     0.05&     0.23&     0.00&     1.00&     0.23&    -0.07&    -0.06&     0.05&    -0.02&    -0.08&         &         &                  \\
(8) Log(Employees)&    12.11&     4.36&     4.48&    24.70&     0.21&     0.08&     0.13&    -0.00&     0.00&    -0.08&    -0.03&         &                  \\
(9) Log(Revenue)&    23.05&     1.44&    18.18&    27.05&     0.63&     0.33&     0.34&    -0.07&    -0.00&    -0.07&    -0.12&     0.32&                  \\
(10) Profit margin&    39.18&    20.66&   -52.87&   100.00&    -0.25&    -0.04&     0.09&     0.02&    -0.01&    -0.01&    -0.01&    -0.16&    -0.27         \\
\hline\hline
\end{tabular}
\end{table}

%% file: logit.tex
\begin{table}[htbp]\centering\footnotesize
\def\sym#1{\ifmmode^{#1}\else\(^{#1}\)\fi}
\caption{Drivers for SBT Adoptions \label{tab:driverSBT}}
\begin{tabular}{l*{1}{S[table-format=2.3]}}
\toprule
                    &\multicolumn{1}{c}{SBT Adoption}\\
\midrule
                    &                   \\
Log(Employees)      &     0.029\sym{*}  \\
                    &   {(0.013)}         \\
\addlinespace
Log(Revenue)        &     0.499\sym{***}\\
                    &   {(0.053)}         \\
\addlinespace
Sales growth        &    -0.478         \\
                    &   {(0.276)}         \\
\addlinespace
Env. score          &     0.038\sym{***}\\
                    &   {(0.004)}         \\
\addlinespace
ROA                 &    -0.008         \\
                    &   {(0.008)}         \\
\addlinespace
Profit margin       &     0.018\sym{***}\\
                    &   {(0.003)}         \\
\addlinespace
Inventory turnover  &    -0.002         \\
                    &   {(0.001)}         \\
\addlinespace
Log(Carbon)         &    -0.190\sym{***}\\
                    &   {(0.030)}         \\
\addlinespace
Constant            &   -16.956\sym{***}\\
                    &   {(1.071)}         \\
\addlinespace
Year fixed effects  &       {Yes}         \\
\midrule
Observations        &      {9091}        \\
Pseudo \(R^{2}\)    &     {0.145}         \\
\bottomrule
\multicolumn{2}{l}{\footnotesize Standard errors in parentheses}\\
\multicolumn{2}{l}{\footnotesize * $ p < .05 $, ** $ p < .01 $, *** $ p < .001 $}\\
\end{tabular}
\end{table}

%% file: h1h2results.tex
\begin{table}[htbp]\centering
\def\sym#1{\ifmmode^{#1}\else\(^{#1}\)\fi}\footnotesize
\caption{Main Results for the Effect of SBTs on Carbon Performance (Hypothesis 1 and 2) \label{tab:resultscarbon}}
\begin{tabular}{l
S[table-format=1.3]
S[table-format=1.3]
S[table-format=1.3]
S[table-format=1.3]
S[table-format=1.3]
}
\toprule
                    &\multicolumn{1}{c}{(1)}&\multicolumn{1}{c}{(2)}&\multicolumn{1}{c}{(3)}&\multicolumn{1}{c}{(4)}&\multicolumn{1}{c}{(5)}\\
                    &\multicolumn{1}{c}{Log(Carbon)}&\multicolumn{1}{c}{Log(Carbon)}&\multicolumn{1}{c}{Log(Carbon)}&\multicolumn{1}{c}{Log(Carbon)}&\multicolumn{1}{c}{Log(Carbon)}\\
\midrule
Treated             &    -1.432\sym{***}&    -3.051\sym{***}&    -3.110\sym{***}&    -3.075\sym{***}&    -2.765\sym{***}\\
                    &   {(0.020)}         &   {(0.884)}         &   {(0.870)}         &   {(0.848)}         &   {(0.779)}         \\
\addlinespace
Post                &     0.090\sym{*}  &     0.089\sym{*}  &                   &     0.079\sym{*}  &     0.059\sym{*}  \\
                    &   {(0.041)}         &   {(0.036)}        &                   &   {(0.035)}         &   {(0.027)}         \\
\addlinespace
Treated $\times$ Post (H1)&    -0.103\sym{*}  &    -0.089         &                   &    -0.089\sym{*}  &    -0.042         \\
                    &   {(0.051)}         &   {(0.045)}         &                   &   {(0.045)}         &   {(0.033)}         \\
\addlinespace
Log(Employees)      &                   &     0.155\sym{*}  &     0.156\sym{*}  &     0.157\sym{*}  &     0.129\sym{*}  \\
                    &                   &   {(0.070)}         &   {(0.069)}         &   {(0.067)}         &   {(0.061)}         \\
\addlinespace
Log(Revenue)        &                   &     0.540\sym{***}&     0.546\sym{***}&     0.558\sym{***}&     0.545\sym{***}\\
                    &                   &   {(0.113)}         &   {(0.113)}         &   {(0.112)}         &   {(0.088)}         \\
\addlinespace
Profit margin       &                   &    -0.005         &    -0.005         &    -0.004         &    -0.004         \\
                    &                   &   {(0.003)}         &   {(0.002)}         &   {(0.002)}         &   {(0.002)}         \\
\addlinespace
Env. score          &                   &     0.003         &     0.003         &     0.003         &     0.002         \\
                    &                   &   {(0.002)}         &   {(0.002)}         &   {(0.002)}         &   {(0.001)}         \\
\addlinespace
Treated $\times$ Post-treatment (t)&                   &                   &    -0.035         &                   &                   \\
                    &                   &                   &   {(0.034)}         &                   &                   \\
\addlinespace
Treated $\times$ Post-treatment (t+1)&                   &                   &    -0.033         &                   &                   \\
                    &                   &                   &   {(0.042)}         &                   &                   \\
\addlinespace
Treated $\times$ Post-treatment (t+2)&                   &                   &    -0.091         &                   &                   \\
                    &                   &                   &   {(0.067)}         &                   &                   \\
\addlinespace
Treated $\times$ Post-treatment (t+3)&                   &                   &    -0.144         &                   &                   \\
                    &                   &                   &   {(0.089)}         &                   &                   \\
\addlinespace
Treated $\times$ Post-treatment (t+4)&                   &                   &    -0.284         &                   &                   \\
                    &                   &                   &   {(0.146)}         &                   &                   \\
\addlinespace
Treated $\times$ Post-treatment (t+5)&                   &                   &    -0.735\sym{**} &                   &                   \\
                    &                   &                   &   {(0.236)}         &                   &                   \\
\addlinespace
Worst in class      &                   &                   &                   &     0.332\sym{*}  &                   \\
                    &                   &                   &                   &   {(0.167)}         &                   \\
\addlinespace
Treated $\times$ Worst in class&                   &                   &                   &    -0.064         &                   \\
                    &                   &                   &                   &   {(0.181)}         &                   \\
\addlinespace
Post $\times$ Worst in class&                   &                   &                   &     0.281         &                   \\
                    &                   &                   &                   &   {(0.311)}         &                   \\
\addlinespace
Treated $\times$ Post $\times$ Worst in class (H2B)&                   &                   &                   &    -0.175         &                   \\
                    &                   &                   &                   &   {(0.315)}         &                   \\
\addlinespace
Best in class       &                   &                   &                   &                   &    -1.233\sym{***}\\
                    &                   &                   &                   &                   &   {(0.246)}         \\
\addlinespace
Treated $\times$ Best in class&                   &                   &                   &                   &     0.704\sym{**} \\
                    &                   &                   &                   &                   &   {(0.270)}         \\
\addlinespace
Post $\times$ Best in class&                   &                   &                   &                   &     0.057         \\
                    &                   &                   &                   &                   &   {(0.127)}         \\
\addlinespace
Treated $\times$ Post $\times$ Best in class (H2A)&                   &                   &                   &                   &    -0.259         \\
                    &                   &                   &                   &                   &   {(0.153)}         \\
\addlinespace
Pre-treatment indicators &        {No}         &        {No}         &       {Yes (Insignificant)}         &        {No}         &        {No}         \\
\addlinespace
Firm fixed effects  &       {Yes}         &       {Yes}         &       {Yes}         &       {Yes}         &       {Yes}         \\
\addlinespace
Year fixed effects  &       {Yes}         &       {Yes}         &       {Yes}         &       {Yes}         &       {Yes}         \\
\midrule
Observations        &      {5256}         &      {5256}         &      {5256}         &     {5256}         &      {5256}         \\
\(R^{2}\)           &     {0.977}         &     {0.980}        &     {0.980}         &     {0.980}         &     {0.984}         \\
\bottomrule
\multicolumn{6}{l}{\footnotesize Standard errors in parentheses, clustered at the firm level}\\
\multicolumn{6}{l}{\footnotesize * $ p<.05 $, ** $ p<.01 $, *** $ p<.001 $}\\
\end{tabular}
\end{table}

%% file: h3results.tex
\begin{table}[htbp]\centering\footnotesize
\def\sym#1{\ifmmode^{#1}\else\(^{#1}\)\fi}
\caption{Main Results for the Effect of SBTs on Non-Carbon Performance (Hypothesis 3) \label{tab:resultsnoncarbon}}
\begin{tabular}{l
S[table-format=1.3]
S[table-format=1.3]
}
\toprule
                    &\multicolumn{1}{c}{(1)}&\multicolumn{1}{c}{(2)}\\
                    &\multicolumn{1}{c}{Env. score}&\multicolumn{1}{c}{Social score}\\
\midrule
Treated             &    -9.436         &   -14.414         \\
                    &  {(12.554)}         &  {(11.409)}         \\
\addlinespace
Post                &     0.551         &    -0.526         \\
                    &   {(0.690)}         &   {(0.698)}         \\
\addlinespace
Treated $\times$ Post (H3)&    -1.065         &    -1.146         \\
                    &   {(0.935)}         &   {(0.971)}         \\
\addlinespace
Log(Employees)      &     1.225         &     2.145\sym{*}  \\
                    &   {(0.982)}         &   {(0.928)}         \\
\addlinespace
Log(Revenue)        &     2.937\sym{*}  &     0.296         \\
                    &   {(1.308)}         &   {(1.247)}         \\
\addlinespace
Profit margin       &    -0.046         &    -0.015         \\
                    &   {(0.045)}         &   {(0.043)}         \\
\addlinespace
Log(Carbon)         &     1.219         &     2.130\sym{**} \\
                    &   {(0.726)}         &   {(0.733)}         \\
\addlinespace
Firm fixed effects  &       {Yes}         &       {Yes}         \\
\addlinespace
Year fixed effects  &       {Yes}         &       {Yes}         \\
\midrule
Observations        &      {5256 }        &      {5256}         \\
\(R^{2}\)           &     {0.835 }        &     {0.827}         \\
\bottomrule
\multicolumn{3}{l}{\footnotesize Standard errors in parentheses, clustered at the firm level}\\
\multicolumn{3}{l}{\footnotesize * $ p<.05 $, ** $ p<.01 $, *** $ p<.001 $}\\
\end{tabular}
\end{table}

%% file: h1h2resultscem.tex
\begin{table}[htbp]\centering
\def\sym#1{\ifmmode^{#1}\else\(^{#1}\)\fi}\footnotesize
\caption{Robustness Test for the Effect of SBTs on Carbon Performance (Coarsened Exact Matching) \label{tab:resultscarboncem}}
\begin{tabular}{l
S[table-format=1.3]
S[table-format=1.3]
S[table-format=1.3]
S[table-format=1.3]
S[table-format=1.3]
}
\toprule
                    &\multicolumn{1}{c}{(1)}&\multicolumn{1}{c}{(2)}&\multicolumn{1}{c}{(3)}&\multicolumn{1}{c}{(4)}&\multicolumn{1}{c}{(5)}\\
                    &\multicolumn{1}{c}{Log(Carbon)}&\multicolumn{1}{c}{Log(Carbon)}&\multicolumn{1}{c}{Log(Carbon)}&\multicolumn{1}{c}{Log(Carbon)}&\multicolumn{1}{c}{Log(Carbon)}\\
\midrule
Treated             &    -1.118\sym{***}&    -5.480\sym{***}&    -5.394\sym{***}&    -5.432\sym{***}&    -5.249\sym{***}\\
                    &   {(0.015)}         &   {(1.179)}         &   {(1.169)}         &   {(1.152)}         &   {(1.185)}         \\
\addlinespace
Post                &     0.079\sym{*}  &     0.064\sym{*}  &                   &     0.057\sym{*}  &     0.062\sym{*}  \\
                    &   {(0.035)}         &   {(0.030)}         &                   &   {(0.028)}         &   {(0.031)}         \\
\addlinespace
Treated $\times$ Post (H1)&    -0.149\sym{**} &    -0.098\sym{*}  &                   &    -0.100\sym{*}  &    -0.056         \\
                    &   {(0.046)}         &   {(0.039)}         &                   &   {(0.040)}         &   {(0.038)}         \\
\addlinespace
Log(Employees)      &                   &     0.477\sym{***}&     0.466\sym{***}&     0.473\sym{***}&     0.456\sym{***}\\
                    &                   &   {(0.118)}         &   {(0.117)}         &   {(0.116)}         &   {(0.119)}         \\
\addlinespace
Log(Revenue)        &                   &     0.303\sym{***}&     0.308\sym{***}&     0.317\sym{***}&     0.339\sym{***}\\
                    &                   &   {(0.089)}         &   {(0.089)}         &   {(0.087)}         &   {(0.084)}         \\
\addlinespace
Profit margin       &                   &    -0.000         &    -0.000         &    -0.000         &    -0.000         \\
                    &                   &   {(0.003)}         &   {(0.003)}         &   {(0.003)}         &   {(0.003)}         \\
\addlinespace
Env. score          &                   &    -0.002\sym{*}  &    -0.002         &    -0.002         &    -0.002         \\
                    &                   &   {(0.001)}         &   {(0.001)}         &   {(0.001)}         &   {(0.001)}         \\
\addlinespace
Treated $\times$ Post-treatment (t)&                   &                   &    -0.060         &                   &                   \\
                    &                   &                   &   {(0.032)}         &                   &                   \\
\addlinespace
Treated $\times$ Post-treatment (t+1)&                   &                   &    -0.073         &                   &                   \\
                    &                   &                   &   {(0.042)}         &                   &                   \\
\addlinespace
Treated $\times$ Post-treatment (t+2)&                   &                   &    -0.112         &                   &                   \\
                    &                   &                   &   {(0.061)}         &                   &                   \\
\addlinespace
Treated $\times$ Post-treatment (t+3)&                   &                   &    -0.092         &                   &                   \\
                    &                   &                   &   {(0.100)}         &                   &                   \\
\addlinespace
Treated $\times$ Post-treatment (t+4)&                   &                   &    -0.309\sym{*}  &                   &                   \\
                    &                   &                   &   {(0.128)}         &                   &                   \\
\addlinespace
Worst in class      &                   &                   &                   &     0.343         &                   \\
                    &                   &                   &                   &   {(0.205)}         &                   \\
\addlinespace
Treated $\times$ Worst in class&                   &                   &                   &    -0.212         &                   \\
                    &                   &                   &                   &   {(0.235)}         &                   \\
\addlinespace
Post $\times$ Worst in class&                   &                   &                   &     0.137         &                   \\
                    &                   &                   &                   &   {(0.165)}         &                   \\
\addlinespace
Treated $\times$ Post $\times$ Worst in class (H2b)&                   &                   &                   &    -0.035         &                   \\
                    &                   &                   &                   &   {(0.174)}         &                   \\
\addlinespace
Best in class       &                   &                   &                   &                   &    -0.517\sym{***}\\
                    &                   &                   &                   &                   &   {(0.138)}         \\
\addlinespace
Treated $\times$ Best in class&                   &                   &                   &                   &     0.179         \\
                    &                   &                   &                   &                   &   {(0.169)}         \\
\addlinespace
Post $\times$ Best in class&                   &                   &                   &                   &    -0.068         \\
                    &                   &                   &                   &                   &   {(0.068)}         \\
\addlinespace
Treated $\times$ Post $\times$ Best in class (H2a)&                   &                   &                   &                   &    -0.362\sym{*}  \\
                    &                   &                   &                   &                   &   {(0.142)}         \\
\addlinespace
Pre-treatment indicators &       {No}       &      {No}        &    {Yes (Insignificant)}      &     {No}         &       {No}        \\
\addlinespace
Year fixed effects  &      {Yes}       &      {Yes}        &      {Yes}        &     {Yes}        &     {Yes}         \\
\addlinespace
Firm fixed effects  &      {Yes}        &      {Yes}       &       {Yes}        &     {Yes}        &     {Yes}         \\
\midrule
Observations        &      {3268}         &     {3268}        &     {3268}         &     {3268}        &      {3268}        \\
\(R^{2}\)           &    {0.977}         &   {0.981}        &   {0.981}        &    {0.981}        &     {0.982}         \\
\bottomrule
\multicolumn{6}{l}{\footnotesize Standard errors in parentheses, clustered at the firm level. Treated$\times$Post-treatment(t+5) is omitted due to the lack of observations.}\\
\multicolumn{6}{l}{\footnotesize * $ p<.05 $, ** $ p<.01 $, *** $ p<.001 $}\\
\end{tabular}
\end{table}

%% file: h1h2resultsaltperc.tex
\begin{table}[htbp]\centering\footnotesize
\def\sym#1{\ifmmode^{#1}\else\(^{#1}\)\fi}
\caption{Alternative Classification for Best- and Worst-in-Class Firms \label{tab:resultscarbonaltperc}}
\begin{tabular}{l
S[table-format=1.3]
S[table-format=1.3]
S[table-format=1.3]
S[table-format=1.3]
}
\toprule
                    &\multicolumn{1}{c}{(1)}&\multicolumn{1}{c}{(2)}&\multicolumn{1}{c}{(3)}&\multicolumn{1}{c}{(4)}\\
                    &\multicolumn{1}{c}{Log(Carbon)}&\multicolumn{1}{c}{Log(Carbon)}&\multicolumn{1}{c}{Log(Carbon)}&\multicolumn{1}{c}{Log(Carbon)}\\
\midrule
Treated             &    -3.109\sym{***}&    -3.413\sym{***}&    -3.007\sym{***}&    -3.055\sym{***}\\
                    &   {(0.806)}         &   {(0.737)}         &   {(0.824)}         &   {(0.715)}         \\
\addlinespace
Post                &     0.074\sym{*}  &     0.069\sym{*}  &     0.076\sym{*}  &     0.075\sym{*}  \\
                    &   {(0.037)}         &   {(0.027)}         &   {(0.037)}         &   {(0.029)}         \\
\addlinespace
Treated $\times$ Post&    -0.086         &    -0.041         &    -0.083         &    -0.049         \\
                    &   {(0.048)}         &   {(0.033)}         &   {(0.049)}         &   {(0.033)}         \\
\addlinespace
Worst in class (20\%)&     0.501\sym{**} &                   &                   &                   \\
                    &   {(0.182)}         &                   &                   &                   \\
\addlinespace
Treated $\times$ Worst in class (20\%)&    -0.194         &                   &                   &                   \\
                    &   {(0.186)}         &                   &                   &                   \\
\addlinespace
Post $\times$ Worst in class (20\%)&     0.025         &                   &                   &                   \\
                    &   {(0.092)}         &                   &                   &                   \\
\addlinespace
Treated $\times$ Post $\times$ Worst in class (20\%)&     0.029         &                   &                   &                   \\
                    &   {(0.105)}         &                   &                   &                   \\
\addlinespace
Log(Employees)      &     0.159\sym{*}  &     0.133\sym{*}  &     0.149\sym{*}  &     0.115         \\
                    &   {(0.064)}         &   {(0.060)}         &   {(0.065)}         &   {(0.060)}         \\
\addlinespace
Log(Revenue)        &     0.581\sym{***}&     0.547\sym{***}&     0.579\sym{***}&     0.593\sym{***}\\
                    &   {(0.110)}         &   {(0.092)}         &   {(0.110)}         &   {(0.100)}         \\
\addlinespace
Profit margin       &    -0.004         &    -0.004         &    -0.003         &    -0.004\sym{*}  \\
                    &   {(0.002)}         &   {(0.002)}         &   {(0.002)}         &   {(0.002)}         \\
\addlinespace
Env. score          &     0.003         &     0.002         &     0.003         &     0.002         \\
                    &   {(0.002)}         &   {(0.001)}         &   {(0.002)}         &   {(0.001)}         \\
\addlinespace
Best in class (20\%)&                   &    -0.780\sym{***}&                   &                   \\
                    &                   &   {(0.169)}         &                   &                   \\
\addlinespace
Treated $\times$ Best in class (20\%)&                   &     0.225         &                   &                   \\
                    &                   &   {(0.194)}         &                   &                   \\
\addlinespace
Post $\times$ Best in class (20\%)&                   &     0.074         &                   &                   \\
                    &                   &   {(0.078)}         &                   &                   \\
\addlinespace
Treated $\times$ Post $\times$ Best in class (20\%)&                   &    -0.246\sym{*}  &                   &                   \\
                    &                   &   {(0.096)}         &                   &                   \\
\addlinespace
Worst in class (25\%)&                   &                   &     0.443\sym{***}&                   \\
                    &                   &                   &   {(0.114)}         &                   \\
\addlinespace
Treated $\times$ Worst in class (25\%)&                   &                   &    -0.155         &                   \\
                    &                   &                   &   {(0.128)}         &                   \\
\addlinespace
Post $\times$ Worst in class (25\%)&                   &                   &     0.050         &                   \\
                    &                   &                   &   {(0.077)}         &                   \\
\addlinespace
Treated $\times$ Post $\times$ Worst in class (25\%)&                   &                   &    -0.001         &                   \\
                    &                   &                   &   {(0.090)}         &                   \\
\addlinespace
Best in class (25\%)&                   &                   &                   &    -0.717\sym{***}\\
                    &                   &                   &                   &   {(0.163)}         \\
\addlinespace
Treated $\times$ Best in class (25\%)&                   &                   &                   &     0.192         \\
                    &                   &                   &                   &   {(0.182)}         \\
\addlinespace
Post $\times$ Best in class (25\%)&                   &                   &                   &     0.023         \\
                    &                   &                   &                   &   {(0.078)}         \\
\addlinespace
Treated $\times$ Post $\times$ Best in class (25\%)&                   &                   &                   &    -0.195\sym{*}  \\
                    &                   &                   &                   &   {(0.091)}        \\
\addlinespace
Year fixed effects  &       {Yes}         &       {Yes}         &       {Yes}        &       {Yes}         \\
\addlinespace
Firm fixed effects  &       {Yes}         &       {Yes}         &       {Yes}         &       {Yes}         \\
\midrule
Observations        &      {5256}         &      {5256}         &      {5256}         &      {5256}         \\
\(R^{2}\)           &     {0.981}         &     {0.984}         &     {0.981}         &     {0.983}         \\
\bottomrule
\multicolumn{5}{l}{\footnotesize Standard errors in parentheses, clustered at the firm level}\\
\multicolumn{5}{l}{\footnotesize * $ p<.05 $, ** $ p<.01 $, *** $ p<.001 $}\\
\end{tabular}
\end{table}

%% file: descriptive_industry.tex
\begin{table}[!h]\centering\footnotesize
\caption{Number of firms by industry \label{tab:desriptive_industry}}
\begin{tabular}{@{}lrrrl@{}}
\toprule
\textbf{Industry}                                 & \textbf{Control} & \textbf{Treated} & \textbf{Total} &  \\ \midrule
Automobiles \& Components                         & 27               & 14               & 41             &  \\
Capital Goods                                     & 59               & 51               & 110            &  \\
Commercial  \& Professional Services              & 9                & 8                & 17             &  \\
Consumer Durables \&   Apparel                    & 12               & 30               & 42             &  \\
Consumer Services                                 & 8                & 8                & 16             &  \\
Energy                                            & 23               & 0                & 23             &  \\
Food \& Staples Retailing                         & 15               & 9                & 24             &  \\
Food, Beverage \& Tobacco                         & 13               & 45               & 58             &  \\
Health Care Equipment \&   Services               & 18               & 5                & 23             &  \\
Household \& Personal   Products                  & 3                & 13               & 16             &  \\
Materials                                         & 49               & 44               & 93             &  \\
Media \& Entertainment                            & 7                & 4                & 11             &  \\
Pharmaceuticals, Biotechnology   \& Life Sciences & 20               & 22               & 42             &  \\
Real Estate                                       & 24               & 12               & 36             &  \\
Retailing                                         & 8                & 15               & 23             &  \\
Semiconductors \&   Semiconductor Equipment       & 13               & 5                & 18             &  \\
Software \& Services                              & 11               & 10               & 21             &  \\
Technology Hardware \&   Equipment                & 15               & 25               & 40             &  \\
Telecommunication Services                        & 7                & 26               & 33             &  \\
Transportation                                    & 19               & 7                & 26             &  \\
Utilities                                         & 17               & 24               & 41             &  \\ \midrule
\textbf{Total}                                    & 377              & 377              & 754            &  \\ \bottomrule
\end{tabular}
\end{table}

%% file: descriptive_country.tex
\begin{table}[!h]\centering\footnotesize
\caption{Number of firms by country \label{tab:desriptive_country}}
\begin{tabular}{@{}lrrr@{}}
\toprule
\textbf{Country}         & \textbf{Control} & \textbf{Treated} & \textbf{Total} \\ \midrule
Argentina                & 1                & 0                & 1              \\
Australia                & 11               & 3                & 14             \\
Austria                  & 4                & 3                & 7              \\
Belgium                  & 4                & 6                & 10             \\
Brazil                   & 10               & 1                & 11             \\
Canada                   & 8                & 6                & 14             \\
Chile                    & 5                & 3                & 8              \\
China                    & 15               & 1                & 16             \\
Colombia                 & 1                & 0                & 1              \\
Cyprus                   & 1                & 0                & 1              \\
Denmark                  & 4                & 7                & 11             \\
Finland                  & 3                & 16               & 19             \\
France                   & 23               & 29               & 52             \\
Germany                  & 12               & 24               & 36             \\
Hong Kong                & 13               & 1                & 14             \\
Hungary                  & 0                & 1                & 1              \\
India                    & 9                & 9                & 18             \\
Indonesia                & 3                & 0                & 3              \\
Ireland                  & 2                & 6                & 8              \\
Israel                   & 1                & 0                & 1              \\
Italy                    & 12               & 10               & 22             \\
Japan                    & 34               & 63               & 97             \\
Korea                    & 18               & 0                & 18             \\
Luxembourg               & 1                & 0                & 1              \\
Macau                    & 1                & 0                & 1              \\
Malaysia                 & 3                & 0                & 3              \\
Mexico                   & 4                & 6                & 10             \\
Netherlands              & 11               & 8                & 19             \\
New Zealand              & 0                & 5                & 5              \\
Norway                   & 2                & 6                & 8              \\
Philippines              & 1                & 0                & 1              \\
Portugal                 & 1                & 1                & 2              \\
Russia                   & 7                & 0                & 7              \\
Singapore                & 3                & 2                & 5              \\
South Africa             & 5                & 2                & 7              \\
Spain                    & 2                & 13               & 15             \\
Sweden                   & 11               & 13               & 24             \\
Switzerland              & 9                & 14               & 23             \\
Taiwan                   & 13               & 8                & 21             \\
Thailand                 & 2                & 0                & 2              \\
Turkey                   & 3                & 1                & 4              \\
United Kingdom           & 22               & 40               & 62             \\
United States of America & 82               & 69               & 151            \\ \midrule
\textbf{Total}           & 377              & 377              & 754            \\ \bottomrule
\end{tabular}
\end{table}

%% file: h1h2results_stddv.tex
\begin{table}[htbp]\centering\footnotesize
\def\sym#1{\ifmmode^{#1}\else\(^{#1}\)\fi}
\caption{Robust test with relative carbon performance measures \label{tab:resultscarbonstddv}}
\begin{tabular}{l
S[table-format=1.3]
S[table-format=1.3]
S[table-format=1.3]
S[table-format=1.3]
}
\toprule
                    &\multicolumn{1}{c}{(1)}&\multicolumn{1}{c}{(2)}&\multicolumn{1}{c}{(3)}&\multicolumn{1}{c}{(4)}\\
                    &\multicolumn{1}{c}{Log(CarbonIntensity)}&\multicolumn{1}{c}{Log(CarbonIntensity)}&\multicolumn{1}{c}{CarbonIndstd}&\multicolumn{1}{c}{CarbonIndstd}\\
\midrule
Treated             &    -3.051\sym{***}&    -3.110\sym{***}&    -0.155         &    -0.213         \\
                    &   {(0.884)}         &   {(0.870)}         &   {(0.479)}         &   {(0.478)}         \\
\addlinespace
Post                &     0.089\sym{*}  &                   &    -0.056         &                   \\
                    &   {(0.036)}        &                   &   {(0.044)}         &                   \\
\addlinespace
Treated $\times$ Post (H1)&    -0.089         &                   &    -0.003         &                   \\
                    &   {(0.045)}         &                   &   {(0.063)}         &                   \\
\addlinespace
Log(Employees)      &     0.155\sym{*}  &     0.156\sym{*}  &     0.052         &     0.052         \\
                    &   {(0.070)}         &   {(0.069)}         &   {(0.039)}         &   {(0.039)}         \\
\addlinespace
Log(Revenue)        &    -0.460\sym{***}&    -0.454\sym{***}&     0.327\sym{***}&     0.323\sym{**} \\
                    &   {(0.113)}         &   {(0.113)}         &   {(0.097)}         &   {(0.099)}         \\
\addlinespace
Profit margin       &    -0.005         &    -0.005         &    -0.004         &    -0.004         \\
                    &   {(0.003)}         &   {(0.002)}         &   {(0.002)}         &   {(0.002)}         \\
\addlinespace
Env. score          &     0.003         &     0.003         &     0.001         &     0.001         \\
                    &   {(0.002)}         &   {(0.002)}         &   {(0.002)}         &   {(0.002)}         \\
\addlinespace
Treated $\times$ Post-treatment (t)&                   &    -0.035         &                   &     0.075         \\
                    &                   &   {(0.034)}         &                   &   {(0.045)}         \\
\addlinespace
Treated $\times$ Post-treatment (t+1)&                   &    -0.033         &                   &     0.065         \\
                    &                   &   {(0.042)}         &                   &   {(0.069)}         \\
\addlinespace
Treated $\times$ Post-treatment (t+2)&                   &    -0.091         &                   &     0.038         \\
                    &                   &   {(0.067)}         &                   &   {(0.102)}         \\
\addlinespace
Treated $\times$ Post-treatment (t+3)&                   &    -0.144         &                   &     0.037         \\
                    &                   &   {(0.089)}         &                   &   {(0.088)}         \\
\addlinespace
Treated $\times$ Post-treatment (t+4)&                   &    -0.284         &                   &     0.114         \\
                    &                   &   {(0.146)}         &                   &   {(0.179)}         \\
\addlinespace
Treated $\times$ Post-treatment (t+5)&                   &    -0.735\sym{**} &                   &    -0.491         \\
                    &                   &   {(0.236)}         &                   &   {(0.326)}         \\
\addlinespace
Pre-treatment indicators &        {No}         &       {Yes}         &        {No}        &       {Yes}         \\
\addlinespace
Firm fixed effects  &       {Yes}         &       {Yes}         &       {Yes}         &       {Yes}         \\
\addlinespace
Year fixed effects  &       {Yes}         &       {Yes}         &       {Yes}         &       {Yes}        \\
\midrule
Observations        &      {5256}         &      {5256}         &      {5256}         &      {5256}         \\
\(R^{2}\)           &     {0.966}         &     {0.967}         &     {0.900}         &     {0.901}         \\
\bottomrule
\multicolumn{5}{l}{\footnotesize Standard errors in parentheses, clustered at the firm level}\\
\multicolumn{5}{l}{\footnotesize * $ p<.05 $, ** $ p<.01 $, *** $ p<.001 $}\\
\end{tabular}
\end{table}

%% file: h1h2results_ratiodv.tex
\begin{table}[htbp]\centering\scriptsize
\def\sym#1{\ifmmode^{#1}\else\(^{#1}\)\fi}
\caption{Robust test with carbon percentage change measures \label{tab:resultscarbonratiodv}}
\begin{tabular}{l
S[table-format=1.3]
S[table-format=1.3]
S[table-format=1.3]
S[table-format=1.3]
}
\toprule
                    &\multicolumn{1}{c}{(1)}&\multicolumn{1}{c}{(2)}&\multicolumn{1}{c}{(3)}&\multicolumn{1}{c}{(4)}\\
                    &\multicolumn{1}{c}{CarbonPercChangeFw}&\multicolumn{1}{c}{CarbonPercChangeFw}&\multicolumn{1}{c}{CarbonPercChangeBk}&\multicolumn{1}{c}{CarbonPercChangeBk}\\
\midrule
Treated             &  -785.624\sym{*}  &  -715.312\sym{*}  &   350.892         &   347.222         \\
                    & {(359.910)}         & {(322.821)}         & {(847.215)}         & {(851.393)}         \\
\addlinespace
Post                &   -36.342         &                   &    12.244         &                   \\
                    &  {(37.583)}         &                   &  {(11.738)}         &                   \\
\addlinespace
Treated $\times$ Post (H1)&    78.314         &                   &   -28.431         &                   \\
                    &  {(89.799)}         &                   &  {(25.917)}         &                   \\
\addlinespace
Log(Employees)      &    48.757         &    47.644         &   -71.681         &   -72.289         \\
                    &  {(48.511)}         &  {(49.188)}         &  {(88.802)}         &  {(89.141)}         \\
\addlinespace
Log(Revenue)        &    51.786         &    51.821         &    66.480         &    67.111         \\
                    &  {(62.730)}         &  {(63.054)}         &  {(39.435)}         &  {(39.065)}         \\
\addlinespace
Profit margin       &    -6.025         &    -6.088         &     0.906         &     0.868         \\
                    &   {(6.343)}         &   {(6.408)}         &   {(2.072)}         &   {(2.062)}         \\
\addlinespace
Env. score          &    -6.747         &    -6.493         &     0.216         &     0.087         \\
                    &   {(7.049)}         &   {(6.763)}         &   {(0.377)}         &   {(0.311)}         \\
\addlinespace
Treated $\times$ Post-treatment (t)&                   &    53.365         &                   &    -4.135         \\
                    &                   &  {(56.461)}         &                   &  {(10.266)}         \\
\addlinespace
Treated $\times$ Post-treatment (t+1)&                   &   -71.618         &                   &     7.525         \\
                    &                   &  {(56.643)}         &                   &  {(16.976)}         \\
\addlinespace
Treated $\times$ Post-treatment (t+2)&                   &   -70.369         &                   &   -18.189         \\
                    &                   &  {(67.010)}         &                   &  {(31.766)}         \\
\addlinespace
Treated $\times$ Post-treatment (t+3)&                   &   -74.701         &                   &   -52.046         \\
                    &                   &  {(65.279)}         &                   &  {(63.659)}         \\
\addlinespace
Treated $\times$ Post-treatment (t+4)&                   &   -73.722         &                   &   -24.412         \\
                    &                   &  {(80.437)}         &                   &  {(33.401)}         \\
\addlinespace
Treated $\times$ Post-treatment (t+5)&                   &  -101.748         &                   &   -24.113         \\
                    &                   & {(126.683)}         &                   &  {(35.482)}         \\
\addlinespace
Pre-treatment indicators &        {No}         &       {Yes}         &        {No}         &       {Yes}         \\
\addlinespace
Firm fixed effects  &       {Yes}         &       {Yes}         &       {Yes}         &       {Yes}         \\
\addlinespace
Year fixed effects  &       {Yes}         &       {Yes}         &       {Yes}         &       {Yes}         \\
\midrule
Observations        &      {4598}         &      {4598}         &      {4916}         &      {4916}         \\
\(R^{2}\)           &     {0.253}         &     {0.255}         &     {0.111}         &     {0.116}         \\
\bottomrule
\multicolumn{5}{l}{\footnotesize Standard errors in parentheses, clustered at the firm level}\\
\multicolumn{5}{l}{\footnotesize * $ p<.05 $, ** $ p<.01 $, *** $ p<.001 $}\\
\end{tabular}
\end{table}